\documentclass[draft,grl]{agu2001}

\usepackage{graphicx}

\authorrunninghead{BEYERLE ET AL.}

\titlerunninghead{GPS OCCULTATION USING ZERO DIFFERENCING}

\authoraddr{G.~Beyerle, S.~Heise, G.~Michalak,
Ch.~Reigber, T.~Schmidt, J.~Wickert,
GeoForschungsZentrum Potsdam (GFZ),
Department~1, Geodesy and Remote Sensing,
Telegrafenberg, D-14473~Potsdam, Germany
(e-mail: gbeyerle@gfz-potsdam.de)}

\begin{document}

\setkeys{Gin}{draft=false}

\title{GPS radio occultation with GRACE:
Atmospheric profiling utilizing the
zero difference technique}

\author{G.~Beyerle,
        T.~Schmidt,
        G.~Michalak,
        S.~Heise,
        J.~Wickert, and
        Ch.~Reigber}
\affil{GeoForschungsZentrum (GFZ) Potsdam, Germany}

\begin{abstract}
Radio occultation events recorded on 28--29~July~2004
by a GPS receiver aboard the \mbox{GRACE-B} satellite are analyzed.
The stability of the receiver clock allows for the derivation of
excess phase profiles using a zero difference technique,
rendering the calibration procedure with concurrent observations
of a reference GPS satellite obsolete.
101~refractivity profiles obtained
by zero differencing and 96~profiles calculated with
an improved single difference method are compared
with co-located ECMWF meteorological analyses.
Good agreement is found at altitudes between 5 and 30~km
with an average fractional refractivity deviation below~1\%
and a standard deviation of 2--3\%.
Results from end-to-end simulations
are consistent with these observations.
\end{abstract}

\begin{article}

\section{Introduction}
\label{se:intro}
In recent years atmospheric sounding by space-based
Global Positioning System (GPS) radio occultation (RO)
is considered a valuable data source for numerical weather
prediction and climate change studies.
From 1995 to 1997 the proof-of-concept GPS/MET mission
performed a series of successful measurement campaigns
\citep{rocken97};
since~2001 a RO instrument operates
aboard the CHAMP (CHAllenging Minisatellite Payload)
\citep{reigber04} satellite.

During an occultation event the RO receiver aboard
the low Earth orbiting (LEO) spacecraft records
dual-frequency signals transmitted from a GPS satellite setting
behind the horizon.
Voltage signal-to-noise ratios ($S\!N\!R_v$) and
carrier phases at the two GPS L-band frequencies
$f_1 = 1.57542$~GHz (L1) and
$f_2 = 1.2276$~GHz (L2) are tracked
with a sampling frequency of typically 50~Hz.
The ionosphere and neutral atmosphere induce
optical path length deviations
from the geometrical distance between transmitter and receiver.
These deviations are denoted as excess phase paths~$\chi$ and
are related to the ray bending
angle~$\alpha$ as a function of impact parameter~$p$.
From $\alpha(p)$ the atmospheric refractivity profile
$N(z) = (n(z)-1)\cdot10^6$ is derived,
where $n(z)$ denotes the real part of the atmospheric
refractive index at altitude~$z$
\citep{kursinski97,hajj02a}.

GPS/MET observations were analyzed with
a double difference method to correct for clock
errors of the GPS transmitters and the receiver
aboard the LEO satellite
\citep{rocken97}.
Since deactivation of Selective Availability (S/A),
an intentional degradation of broadcast ephemeris data,
on 2~May~2000, the GPS clock errors are reduced by orders of magnitude.
Without S/A GPS clocks are sufficiently stable to replace
double differencing by the single difference technique
thereby eliminating the need for concurrent high-rate
ground station observations
\citep{wickert02}.

On 28--29~July 2004 the GPS RO receiver aboard the \mbox{GRACE-B} satellite,
the second of the two GRACE satellites
\citep{tapley04a},
was activated for 25~hours to test the receiver's occultation mode
\citep{wickert05a}.
The \mbox{GRACE-B} receiver clock time was adjusted to
GPS coordinate time only once during the measurement period.
Thus, the \mbox{GRACE-B} clock solutions can be interpolated
from the precise orbit
determination's (POD) temporal resolution of 30~s to the occultation receiver's
resolution of 20~ms and the excess phase path is obtained without simultaneous
observations of a reference GPS satellite (zero differencing).

\section{Methodology and Data Analysis}
\label{se:method}
The L1 and L2 signals transmitted by the occulting
GPS satellite are recorded by the receiver's signal tracking
phase-locked loops.
Their numerically-controlled oscillators (NCOs) are
phase-locked to the signal carriers transmitted by the GPS satellite
which in turn are phase-locked to the GPS transmitter clock.
We may regard the NCOs as clocks
that are synchronized to the transmitter clock of the occulting
satellite.
The occulting GPS clock time, as recorded by the receiver NCO,
is denoted by $\hat{t}^O_k$.
($k=1$ for L1, $k=2$ for L2; though, in the following the subscript~$k$
is omitted.)
During an occultation event the receiver measures both
$\hat{t}^O$ and the corresponding LEO clock time $\hat{t}^L$.

\subsection{Zero and Single Differencing Methods}
\label{ss:zerosingle}

Since the GPS and LEO satellites move within Earth's gravitational potential,
the proper time~$\bar{t}$, the coordinate time~$t$ and the clock time~$\hat{t}$
in general will not agree.
In order to extract the excess phase paths $\chi(t)$ from the
measurements~$L^{OL} \equiv c\,(\hat{t}^L - \hat{t}^O)$,
the LEO clock corrections $(\hat{t}^L - {t}^L)$ and
the occulting GPS clock corrections $(\hat{t}^O - {t}^O)$
have to be taken into account
\citep{hajj02a}.
Here, $c$ denotes the velocity of light in vacuum.
We note that both, $(\hat{t}^L - {t}^L)$ and $(\hat{t}^O - {t}^O)$,
include relativistic effects relating coordinate time to proper time.
Thus, the zero differencing (ZD) equation is
\begin{eqnarray}
\label{eq:zerodiff}
{t}^L - {t}^O = \frac{1}{c}\,L^{OL}
                - (\hat{t}^L - {t}^L)
                - ({t}^O - \hat{t}^O) \; .
\end{eqnarray}
The left hand side of Eqn.~\ref{eq:zerodiff} contains
the excess phase path~$\chi$ through
\begin{eqnarray}
\label{eq:loterms}
{t}^L - {t}^O &\approx& \frac{1}{c}\,\left(d^{OL}+\chi\right)
\end{eqnarray}
where contributions from the gravitational time delay are neglected
\citep{hajj02a} and
$d^{OL} \equiv |\vec{r}^O-\vec{r}^L|$.
The positions of the occulting GPS and the LEO satellite,
$\vec{r}^{\,O}$ and $\vec{r}^{\,L}$,
and their clock corrections,
$(\hat{t}^O - {t}^O)$
and
$(\hat{t}^L - {t}^L)$,
are obtained from the POD
\citep{koenig04}.
CHAMP's and \mbox{GRACE-B}'s orbits and clock solutions are provided
at a temporal resolution of 30~s.

GRACE-B's receiver clock is adjusted to GPS coordinate time
only once
during the 25~hour measurement period on 28~July 16:24~UTC.
The difference between clock time
and coordinate time, $(\hat{t}^L - {t}^L)$, at the
required temporal resolution of 20~ms
could therefore successfully be interpolated
from the 30~s clock solutions.
The receiver clock aboard CHAMP, on the other hand, is
adjusted once per second
to achieve a 1~$\mu$s maximum deviation with respect to coordinate time.
(See discussion of Fig.~\ref{fg:clkerrchampgrace} below.)
These adjustments introduce discontinuities that cannot
be corrected for on the basis of the 30~s clock solutions and
require the simultaneous observation of a referencing GPS satellite
(single differencing (SD))
\begin{eqnarray}
\label{eq:singlediff}
t^L - t^O
  &=& \frac{1}{c}\,(L^{OL}-L^{RL}_c)
     + ({t}^L     - {t}^R)
     + ({t}^R     - \hat{t}^R)
     - ({t}^O     - \hat{t}^O)
\end{eqnarray}
with superscript~$R$ indicating the reference satellite
\citep{wickert02}.
The reference link observation corrected for
ionospheric signal propagation, $L^{RL}_c$, is calculated from
\begin{eqnarray}
\label{eq:ionocorr1}
L^{RL}_c \equiv c_1\,L^{RL}_{L1} - c_2\,L^{RL}_{L2} \; .
\end{eqnarray}
with parameters $c_1$ and $c_2$ defined as
$c_i = (f_i)^2/[(f_1)^2 - (f_2)^2]$, $i=1,2$.

GFZ's operational RO processing system
\citep{wickert04a}
improves the ionospheric correction of the reference link observations
(Eqn.~\ref{eq:ionocorr1})
by assuming that quiet ionospheric conditions prevail
(modified single differencing, in the following abbreviated as SDM).
Within the SDM approach $L^{RL}_c$ is replaced by
\begin{eqnarray}
\label{eq:ionocorr2}
\tilde{L}^{RL}_c \equiv L^{RL}_{L1} + c_2\,\langle L^{RL}_{L1} - L^{RL}_{L2} \rangle
\end{eqnarray}
where $\langle \cdot \rangle$ denotes a smoothing (low-pass) filter.
Since the noise level of $L^{RL}_{L2}$ is much larger than
the noise level of $L^{RL}_{L1}$ (see Fig.~\ref{fg:looperr})
the filter effectively reduces the L2 noise.

\subsection{Comparison with ECMWF}
\label{ss:ecmwf}

The observed GRACE refractivity profiles are intercompared with
meteorological analysis results provided by the
European Centre for Medium-Range Weather Forecasts (ECMWF).
Pressure and temperature values from the
ECMWF fields are calculated by linear
interpolation between grid points (0.5$^\circ\times0.5^\circ$ resolution).
Linear interpolation in time is performed between
6~h analyses fields.
The comparison between RO observation and meteorological analysis
is performed on the 60~ECMWF model levels
ranging from the ground surface up to 0.1~hPa (about 60~km altitude)
after converting the pressure levels to height coordinates.
Vertical spacing of the model grid points increases from about 200~m
at 1~km altitude to about 700~m at 10~km altitude.

\subsection{Simulation Study}
\label{ss:simul}

We interpret the \mbox{GRACE-B} observations using results from end-to-end
simulations.
The refractivity field is assumed to be spherically symmetric and
calculated from the corresponding ECWMF profile;
electron densities are modelled using the
Parameterized Ionospheric Model (PIM)
\citep{daniell95}.
The occultation events are simulated with circular
approximations of the true
\mbox{GRACE-B} and GPS satellite orbits.
Signal amplitudes and carrier phases on the occultation link are
calculated from bending angles with the inverse CT technique
\citep{gorbunov03a}
modified for circular orbits (inverse FSI or CT2 method).
The amplitude is normalized to the observation
averaged over the first 10~s of the occultation.
The reference link data are assumed to be multipath-free and the
carrier phases are determined using geometric optics, their signal
amplitudes are taken to be constant.
Again, their magnitude is extracted from the observation.
On both links ray bending due to the ionosphere and neutral atmosphere
is taken into account.

In principle, carrier tracking loop errors as well as clock
instabilities and ionospheric disturbances contribute
to the uncertainties of the observations
$L^{OL}$ and $L^{RL}$.
For a given amplitude $S\!N\!R_v$
the tracking loop phase noise~$\sigma_c$ can be estimated from
\citep{ward96}
\begin{eqnarray}
\label{eq:phasenoise}
\sigma_c = \frac{\lambda}{2\pi}\,
  \sqrt{\frac{2\,B_w\,1\,\textrm{s}}{(S\!N\!R_v)^2}\,
  \left(1+\frac{1\,\textrm{s}}{(S\!N\!R_v)^2\,T}\right)}
\end{eqnarray}
with carrier wavelength~$\lambda$, sampling time~$T$
and carrier loop bandwidth~$B_w$.

\begin{figure}
\noindent
\includegraphics[width=20pc,keepaspectratio]
{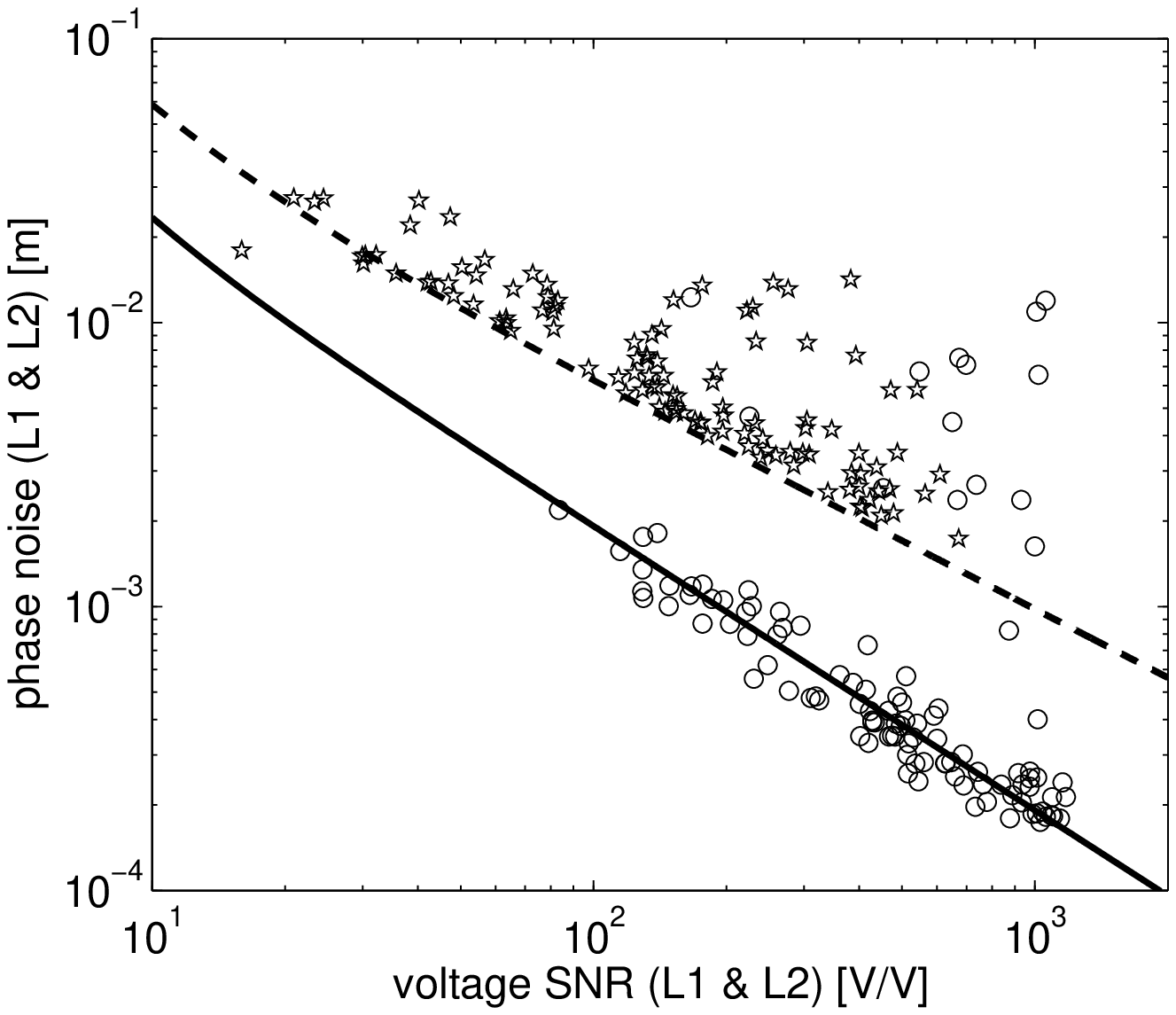}
\caption{
\label{fg:looperr}
Carrier phase noise
extracted from the reference link observations as a function
of voltage $S\!N\!R_v$.
The L1 and L2 observations extracted from 109~reference
link profiles are marked as circles and stars, respectively.
The L1~phase noise (full line) is calculated from Eqn.~\ref{eq:phasenoise}
with $T=20$~ms and $B_w=20$~Hz;
the corresponding L2 noise (dashed line)
is a heuristic parameterization since
additional losses from semi-codeless or codeless L2 tracking
are not taken into account by Eqn.~\ref{eq:phasenoise}.
}
\end{figure}

The simulated L1~phase noise is calculated from Eqn.~\ref{eq:phasenoise}
using $T=20$~ms and $B_w=20$~Hz,
since this parametrization is in approximate agreement with the observed
$L^{RL}_{L1}$ noise levels which are plotted in
Fig.~\ref{fg:looperr} (circles).
The full line marks the result from Eqn.~\ref{eq:phasenoise}.
The corresponding L2 noise values, however, estimated from
Eqn.~\ref{eq:phasenoise} significantly underestimate the observations
since additional losses from semi-codeless or codeless L2 tracking
are not taken into account by Eqn.~\ref{eq:phasenoise}.
A heuristic parametrization is therefore used to simulate
the L2 phase noise (dashed line in Fig.~\ref{fg:looperr}).

The uncertainties of both observations,
$L^{RL}_{L1}$ and $L^{RL}_{L2}$,
exhibit a distinct dependency on $S\!N\!R_v$
(Fig.~\ref{fg:looperr}).
Since the clock error is expected to be independent of $S\!N\!R_v$
we conclude from Fig.~\ref{fg:looperr}
that tracking loop errors dominate the clock uncertainties.
Enhanced noise values in some of the observations are presumably attributed
to ionospheric disturbances.
From the simulated $S\!N\!R_v(t)$ and $\chi(t)$ data
bending angle profiles are reconstructed using the FSI method
\citep{jensen03}
and (above 20~km altitude) using the geometric optics method.
Finally, refractivity profiles are retrieved by Abel-transforming
\citep{fjeldbo71}
the bending angle profiles thereby closing the simulation loop.

\section{Discussion}
\label{se:disc}
Between 6:03~UTC on 28~July 2004 and 7:09~UTC on 29~July 2004
the GPS receiver aboard \mbox{GRACE-B} was activated to test occultation
measurement mode.
During these 25~hours the receiver recorded 109~occultation events
lasting longer than~40~s;
101~events could be successfully
converted to atmospheric refractivity profiles with ZD and SDM processing,
8~were removed as outliers.
A profile is regarded as outlier if the fractional refractivity error
with respect to ECMWF exceeds~10\% at altitudes above~10~km.
The corresponding yield for SD is 96~profiles.

\begin{figure}
\noindent
\includegraphics[width=20pc,keepaspectratio]
{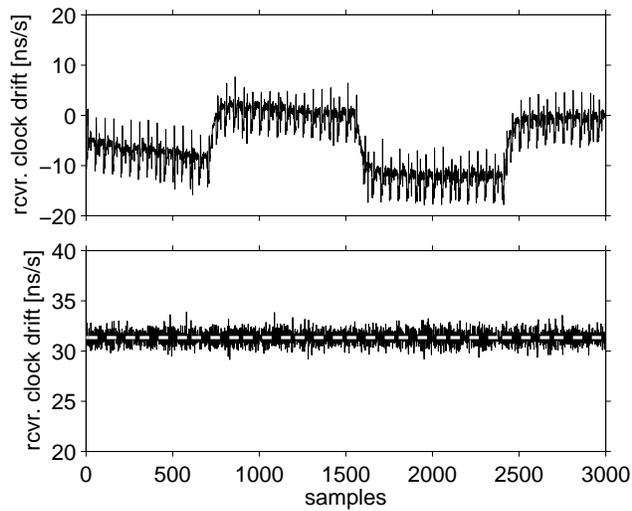}
\caption{
\label{fg:clkerrchampgrace}
Receiver clock drift derived from \mbox{GRACE-B} and CHAMP reference link data.
Top: receiver clock drift observed in a CHAMP occultation event
on~28~July 2004, 0:18~UTC at~69.4$^\circ$N, 12.5$^\circ$W.
Bottom: clock drift observed in a GRACE occultation event
on~28~July 2004, 6:41~UTC at~66.6$^\circ$S, 13.9$^\circ$E (black).
Note the change in vertical scale.
}
\end{figure}

From the reference link observations~$L^{RL}$
the difference between receiver clock time and proper time,
$(\hat{t}^L-\bar{t}^L)$,
can be estimated assuming quiet ionospheric conditions
\citep{wickert02}.
Its temporal evolution,
$\Delta(\hat{t}^L-\bar{t}^L)/T$,
extracted from the first \mbox{GRACE-B} occultation measurement
on 28~July~2004, 6:10~UTC at 55.3$^\circ$N, 22.3$^\circ$E,
is plotted in Fig.~\ref{fg:clkerrchampgrace}, bottom panel, black line.
Here,
$\Delta(x_n) \equiv x_{n+1} - x_n$ denotes
the forward difference operator and the sampling rate
is $T = 20$~ms.
The mean clock drift of $31.3347\pm0.6305$~ns/s
is consistent with a $31.3338$~ns/s mean drift
obtained from \mbox{GRACE-B} precise orbit calculations (white line).

For comparison the top panel of Fig.~\ref{fg:clkerrchampgrace}
shows an occultation event recorded by CHAMP on the same day,
0:18~UTC at~69.4$^\circ$N, 12.5$^\circ$W.
The CHAMP clock drift
($-1.0642\pm8.5187$~ns/s)
exhibits periodic discontinuities
of about~$10$~ns/s about every~18~s;
in addition, the CHAMP clock is adjusted
every second
in order to meet the design specification of 1~$\mu$s absolute time signal
error with respect to coordinate time.
These discontinuities cannot be modelled from the 30~s POD
data eliminating at present the possibility of
zero differencing with CHAMP.
Since the data sets shown in Fig.~\ref{fg:clkerrchampgrace}
are extracted from reference link observations,
the random components contain contributions from the ionosphere
as well as the tracking loops and should not be interpreted as clock noise.

\begin{figure}
\noindent
\includegraphics[width=20pc,keepaspectratio]
{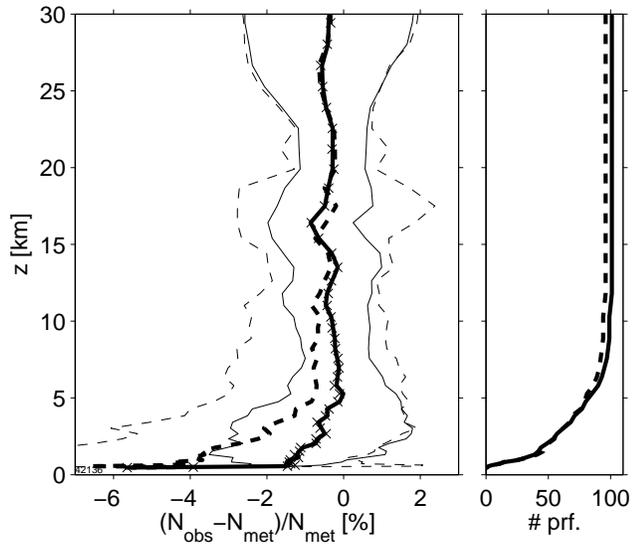}
\caption{
\label{fg:frcnstat}
Left: Mean fractional refractivity
deviation between GRACE observations and ECMWF analyses.
The data set comprises 101~zero differencing (solid line)
and 96~single differencing (dashed line) results.
Thin lines indicate the 1-$\sigma$
standard deviations.
Retrieval results from modified single differencing
are marked as crosses.
Right: number of data points retrieved as a function
of altitude.
}
\end{figure}

The mean fractional refractivity error
$\langle (N_{obs}-N_{met})/N_{met}\rangle$
derived from the \mbox{GRACE-B} observations is plotted in
Fig.~\ref{fg:frcnstat} (left panel),
where $N_{obs}$ and $N_{met}$ denote the observed and ECMWF
refractivities, respectively.
The 1-$\sigma$ standard deviations
are marked as thin lines.
The number of extracted data points as a function
of altitude is plotted in the right panel.

The ZD results are characterized by smaller standard deviations
at altitudes below 20~km since noise contributions from the
reference link are present in the SD, but not in the ZD excess phases.
Above 20~km multipath is assumed to be negligible and
bending angles are calculated using geometric optics.
In addition, the SDM results are marked by crosses
in the left panel of Fig.~\ref{fg:frcnstat}.
The almost perfect agreement with the ZD profiles,
both in terms of bias (crosses) and standard deviation (not shown),
emphasizes the impact of the respective ionospheric correction
procedure (Eqn.~\ref{eq:ionocorr1} vs.\ Eqn.~\ref{eq:ionocorr2})
on the tropospheric retrieval results.

Fig.~\ref{fg:frcnstat} suggests that ZD processing reduces the
lower tropospheric refractivity bias compared to the SD results.
At low $S\!N\!R_v$ substantial phase noise contributions
are present in the excess phase path profile~$\chi(t)$
thereby degrading the radioholographic interference patterns
encoded in $\chi(t)$.
These patterns originate from interfering rays associated with
large negative vertical refractivity gradients that are caused
by humidity layers in the lower troposphere.
From the degraded interference patterns the FSI algorithm
underestimates the large bending angles produced by the
vertical refractivity gradients leading to smaller refractivities
and, ultimately, a negative bias with respect to ECMWF.

\begin{figure}
\noindent
\includegraphics[width=20pc,keepaspectratio]
{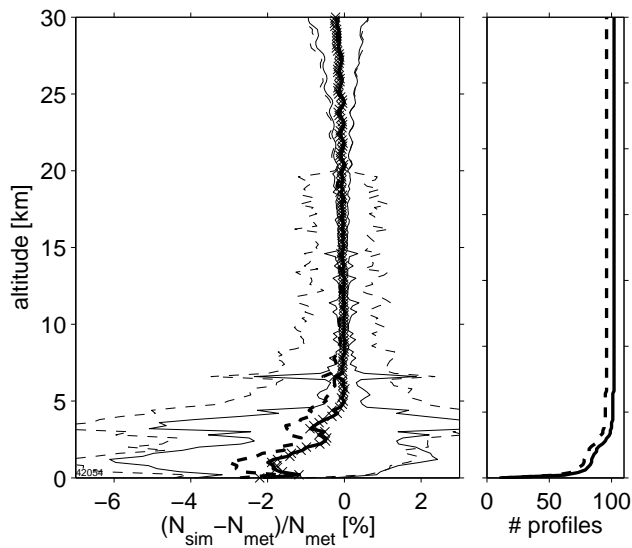}
\caption{ \label{fg:frcnstatsim}
Same as Fig.~\ref{fg:frcnstat}, except that results from end-to-end
simulations are shown.
Left:
the mean error for ZD and SD processing
are plotted as thick solid and dashed lines, respectively.
Thin lines indicate the 1-$\sigma$ standard deviations.
The result obtained from a SD simulation
with noise-free reference link phase observations (crosses)
is in almost perfect agreement with the ZD result.
Right: number of data points retrieved as a function
of altitude.
}
\end{figure}

We substantiate our interpretation of the
difference between ZD and SD data with simulation results
shown in Fig.~\ref{fg:frcnstatsim}.
We note that the non-zero mean fractional refractivity error
and standard deviation is caused by including
L1/L2 carrier phase noise within the simulation.
Above 20~km altitude the bending angles derived from
the FSI are replaced by the values obtained from
the geometric optics method, since the occurrence
of multipath ray propagation in the stratosphere can be excluded.
The simulations reproduce the over-all differences
between the observed ZD and SD refractivities (Fig.~\ref{fg:frcnstat})
as well as an (albeit smaller) bias between the two data sets
at altitudes below 10~km.
An additional SD simulation was conducted with
the noise level of the reference link observations
reduced to zero.
Since the result (marked by crosses in Fig.~\ref{fg:frcnstatsim})
is in excellent agreement with the ZD profile throughout the full
altitude range,
the observed bias between the ZD and SD results in the lower troposphere
is evidently caused by the reference link phase noise.

\section{Conclusion}
\label{se:concl}
First radio occultation events observed by the \mbox{GRACE-B} satellite
are successfully analyzed using zero differencing.
In the upper troposphere and lower stratosphere
the refractivities derived from zero and single differencing
are in good agreement with the corresponding
ECMWF meteorological analyses.
Zero differencing allows for a reduction in the amount of down-link data.
More significantly, zero differencing reduces the noise level on the excess
phase paths and thereby yields less-biased refractivities within regions of
multipath signal propagation in the lower troposphere compared to single
differencing.
However, for quiet ionospheric conditions a modified single differencing
method can be applied producing refractivities that are in excellent
agreement with the zero differencing results.

\begin{acknowledgments}
Help and support from
F.~Flechtner, L.~Grunwaldt, W.~K\"ohler, and F.-H.~Massmann
are gratefully acknowledged.
Comments by anonymous reviewers on earlier versions
improved this paper considerably.
We thank JPL for providing the GRACE occultation raw data.
The German Ministry of Education and Research
(BMBF) supports the GRACE project within the GEOTECHNOLOGIEN
geoscientific R+D program under grant~03F0326A.
The European Centre for Medium-Range Weather Forecasts
provided meteorological analysis fields.
\end{acknowledgments}

\end{article}
\newpage

\begin{thebibliography}{14}
\providecommand{\natexlab}[1]{#1}
\expandafter\ifx\csname urlstyle\endcsname\relax
  \providecommand{\doi}[1]{doi:\discretionary{}{}{}#1}\else
  \providecommand{\doi}{doi:\discretionary{}{}{}\begingroup
  \urlstyle{rm}\Url}\fi

\bibitem[{\textit{Daniell et~al.}(1995)\textit{Daniell, Brown, Anderson, Fox,
  Doherty, Decker, Sojka, and Schunk}}]{daniell95}
Daniell, R.~E. et~al. (1995), Parameterized ionospheric
  model: {A} global ionospheric parameterization based on first principles
  models, \textit{Radio Sci.}, \textit{30}(5), 1499--1510.

\bibitem[{\textit{Fjeldbo et~al.}(1971)\textit{Fjeldbo, Kliore, and
  Eshleman}}]{fjeldbo71}
Fjeldbo, G., A.~J. Kliore, and V.~R. Eshleman (1971), The neutral atmosphere of
  {V}enus as studied with the {M}ariner {V} radio occultation experiments,
  \textit{Astron. J.}, \textit{76}(2), 123--140.

\bibitem[{\textit{Gorbunov}(2003)}]{gorbunov03a}
Gorbunov, M.~E. (2003), An asymptotic method of modeling radio occultations,
  \textit{J. Atmos. Solar-Terr. Phys.}, \textit{65}, 1361--1367,
  \doi{10.1016/j.jastp.2003.09.001}.

\bibitem[{\textit{Hajj et~al.}(2002)\textit{Hajj, Kursinski, Romans, Bertiger,
  and Leroy}}]{hajj02a}
Hajj, G.~A., et~al.
  (2002), A technical description of atmospheric sounding by {GPS} occultation,
  \textit{J. Atmos. Solar-Terr. Phys.}, \textit{64}(4), 451--469.

\bibitem[{\textit{Jensen et~al.}(2003)\textit{Jensen, Lohmann, Benzon, and
  Nielsen}}]{jensen03}
Jensen, A.~S. et~al. (2003), Full spectrum
  inversion of radio occultation signals, \textit{Radio Sci.}, \textit{38}(3),
  1040, \doi{10.1029/2002RS002763}.

\bibitem[{\textit{K\"onig et~al.}(2004)\textit{K\"onig, Michalak, Neumayer,
  Schmidt, Zhu, Meixner, and Reigber}}]{koenig04}
K\"onig, R., et~al. (2004), Recent developments in {CHAMP} orbit determination at
  {GFZ}, in \textit{Earth Observation with CHAMP, Results from Three Years in
  Orbit}, edited by
  C.~Reigber, et~al., pp. 65--70, Springer,
  Berlin.

\bibitem[{\textit{Kursinski et~al.}(1997)\textit{Kursinski, Hajj, Schofield,
  Linfield, and Hardy}}]{kursinski97}
Kursinski, E.~R., et~al.
  (1997), Observing {E}arth's atmosphere with radio occultation measurements
  using {G}lobal {P}ositioning {S}ystem, \textit{J. Geophys. Res.},
  \textit{19}(D19), 23,429--23,465.

\bibitem[{\textit{Reigber et~al.}(2004)\textit{Reigber, L\"uhr, Schwintzer, and
  Wickert}}]{reigber04}
Reigber, C., et~al. (2004), \textit{Earth
  Observation with CHAMP: {R}esults from Three Years in Orbit},
  Springer--Verlag, Berlin Heidelberg New York.

\bibitem[{\textit{Rocken et~al.}(1997)}]{rocken97}
Rocken, C., et~al. (1997), Analysis and validation of {GPS/MET} data in the
  neutral atmosphere, \textit{J. Geophys. Res.}, \textit{102}(D25),
  29,849--29,866.

\bibitem[{\textit{Tapley et~al.}(2004)\textit{Tapley, Bettadpur, Watkins, and
  Reigber}}]{tapley04a}
Tapley, B.~D. et~al. (2004), The gravity
  recovery and climate experiment: {M}ission overview and early results,
  \textit{Geophys. Res. Lett.}, \textit{31}, L09607,
  \doi{10.1029/2004GL019920}.

\bibitem[{\textit{Ward}(1996)}]{ward96}
Ward, P. (1996), \textit{Understanding GPS: Principles and applications (edited
  by E.~D.~Kaplan)}, chap. Satellite signal acquisition and tracking, Artech
  House, Boston, London.

\bibitem[{\textit{Wickert et~al.}(2002)\textit{Wickert, Beyerle, Hajj,
  Schwieger, and Reigber}}]{wickert02}
Wickert, J., et~al. (2002), {GPS}
  radio occultation with champ: Atmospheric profiling utilizing the space-based
  single difference technique, \textit{Geophys. Res. Lett.}, \textit{29}(8),
  1187, \doi{10.1029/2001GL013982}.

\bibitem[{\textit{Wickert et~al.}(2004)\textit{Wickert, Schmidt, Beyerle,
  K\"onig, Reigber, and Jakowski}}]{wickert04a}
Wickert, J., et~al.
  (2004), The radio occultation experiment aboard {CHAMP}: {O}perational data
  processing and validation of atmospheric parameters, \textit{J. Meteorol.
  Soc. Jpn.}, \textit{82}(1B), 381--395.

\bibitem[{\textit{Wickert et~al.}(2005)\textit{Wickert, Beyerle, K\"onig,
  Grunwaldt, Heise, Michalak, Reigber, and Schmidt}}]{wickert05a}
Wickert, J., et~al. (2005), {GPS} radio occultation with {CHAMP} and
  {GRACE}: {A} first look at a new and promising satellite configuration for
  global atmospheric sounding, \textit{Ann. Geophys.}, \textit{23}, 653--658.

\end{thebibliography}
\end{document}